\documentclass[twocolumn]{aa}

\usepackage{xcolor}
\usepackage{graphicx}
\usepackage{longtable}
\usepackage{amsmath}
\usepackage{array}
\usepackage{threeparttable}
\newcolumntype{C}[1]{>{\centering\arraybackslash}m{#1}}

\usepackage{txfonts}
\begin{document}

   \title{Peering into the heart of 3CR radio galaxies}

   \subtitle{A very long baseline interferometry perspective on optical-radio classifications at parsec scales}

   \author{P. Grandi
          \inst{1}
          \and
          G. Giovannini\inst{2,3}
          \and 
          E. Torresi\inst{1}
          \and
          B. Boccardi\inst{4}
          }

  \institute{INAF-Osservatorio di Astrofisica e Scienza dello Spazio di Bologna, Area della Ricerca CNR, Via Gobetti 101, I-40129 Bologna, Italy\\
              \email{paola.grandi@inaf.it}
         \and
        INAF – Istituto di Radioastronomia, via Gobetti 101, 40129 Bologna, Italy
        \and
        Dipartimento di Fisica e Astronomia – University of Bologna via Gobetti 93/2, Italy 
        \and
        Max-Planck-Institut f\"{u}r Radioastronomie, Auf dem H\"{u}gel 69, D-53121 Bonn, Germany}


   \date{Received November 15, 2024; accepted May 25, 2025}

 
  \abstract{
 Historically, luminous edge-brightened Fanaroff-Riley type II (FRII) radio galaxies have been associated with radiatively efficient cold accretion disks. In contrast, faint edge-darkened Fanaroff-Riley Type Is (FRIs) are sustained by hot, inefficient accretion flows.
However, several sources deviate from this paradigm, showing FRII morphologies and radiatively inefficient accretion.

Three hypotheses have been proposed to explain the observations: (1)  The evolutionary scenario: initially strong accretor FRIIs switch to having a hot inefficient flow due to the depletion of available material.  (2) The environment scenario: radio structures are mainly shaped by the environment.\ (3) The engine scenario: both radiatively efficient (FRII-HERGs) and inefficient flows (FRII-LERGs) can expel jets powerful enough to maintain collimation up to kiloparsec distances from the nucleus. 

We explored these scenarios by studying the jet properties of 3CR radio galaxies using very long baseline interferometry data from the Fundamental Radio Catalog   
 to investigate the distinction between FRII-LERGs and FRII-HERGs. In particular, we analyzed the 8 GHz luminosity ($L_{8GHz}$) distribution across different optical-radio classes at milliarcsecond scales. Our results favor the engine scenario. The separation between different radio-optical classes is evident even on parsec scales. In particular, the separation between FRII classes with different accretion rates occurs near the central engine before the jets reach kiloparsec-scale distances.}

   {}

   \keywords{Galaxies: active -- Galaxies: jets}

   \maketitle
%

\section{Introduction}

Jetted active galactic nuclei  have central engines that differ in their efficiency in converting gravitational potential energy into radiation and kinetic energy.
Radio galaxies (RGs), a type of jetted active galactic nucleus, are affected by moderate relativistic effects because the jet is not aligned with the observer's line of sight. This favorable orientation makes them ideal targets for analyzing the relationship between accretion and ejection at different wavelengths. In RGs, the nuclear region, which is not overwhelmed by nonthermal boosted jet radiation, is directly accessible for study, and the ejected plasma, which is mildly affected by projection effects, can be explored across various spatial scales.

Radio galaxies are divided into two groups: Fanaroff-Riley Type I (FRI) and Fanaroff-Riley Type II (FRII; \citealt{fanaroff74}).
FRI RGs are characterized by collimated radio jets that decrease in brightness as they move away from the central nucleus. In these sources, the radio emission is brightest near the core, with the jets dispersing into broader, fainter radio structures at larger distances. In contrast, FRII RGs have jets that remain collimated over greater distances, terminating in bright radio lobes at the outer edges of the source, often with distinct hotspots marking the interaction between the jets and the surrounding medium.

From an optical perspective, RGs are separated into high-excitation radio galaxies (HERGs) and low-excitation radio galaxies \citep[LERGs;][]{JaR}. HERGs are further subclassified as broad-line radio galaxies (BLRGs) if they exhibit broad lines in their optical spectra or narrow-line radio galaxies (NLRGs) if they do not. In BLRGs, the broad-line region is directly observable (i.e., the jet inclination angle is less than 30-40 degrees). In contrast, in NLRGs, the angle between the jet and the line of sight is larger, and the torus obscures the more internal regions. From here on, we refer to HERGs generally without specifying the subclass.
Operatively, the optical classification relies on inspecting the emission lines from the narrow-line region: the intensities of high- and low-excitation lines are compared to provide insights into the ionization state of the gas. As the measured line ratios indicate the central source's ability to photo-ionize the circumnuclear gas, the distinction between LERGs and HERGs reflects differences in the accretion regime. 
HERGs convert most of the gravitational power into radiation through a geometrically thin, optically thick disk, the Shakura-Sunyaev disk (SS-disk; \citealt{ss}), and provide many photoionizing photons. In LERGs, the less efficient process of producing radiation (the radiative inefficient accretion flow) reduces the impact of nuclear emissions on the narrow-line regions. Different models have been proposed to describe the radiative inefficient accretion flow. One of the most suggestive is the advection-dominated accretion flow, which assumes a hot flow that is geometrically thick but optically thin. The decoupling between electrons and ions in the accreting matter prevents an efficient radiative cooling of the accreting matter \citep{yuan}. The viscously dissipated energy is stored in the gas and advected into the black hole, rather than being radiated.
Triggered by the millimeter high-spatial-resolution observations of M87  with the Event Horizon Telescope \citep{EHT}, \cite{globus} proposed a new interpretative framework.
In their model,  the rotational energy of the black hole, rather than the gravitational power, is the primary energy source for both the jets and the disk. Material accreting at the Bondi radius\footnote{The Bondi radius describes the black hole's sphere of influence and is defined as $R_\text{Bondi} = \frac{2 G M_\text{BH}}{c_{\rm s}^2}$. Here $c_{\rm s}$ is the sound speed, which depends on the density and temperature of the surrounding medium \citep{bondi52}.} is first deposited in a cold disk and then expelled through a magneto-centrifugal wind. The same wind is then used to collimate the jet produced in the innermost regions. 

Historically, RGs have been divided into two groups: FRI-LERGs and FRII-HERGs. This one-to-one separation implicitly assumes that jets are more potent if the central engine hosts an SS-disk.
Although appealing, this classification is oversimplified.
For instance, a study of 3CR radio galaxies \citep{3CR} up to z<0.3 \citep{buttiglione09,buttiglione10,buttiglione11} indicates that approximately 25\% of the RGs are classified as FRII-LERGs, i.e., they display a bright-edge radio morphology but inefficient accretion. An X-ray study of 79 3CR RGs with optical and radio classifications has unequivocally shown that FRII-LERGs exhibit properties between those of FRI-LERGs and FRII-HERGs in terms of 2-10 keV luminosity and intrinsic ($N_\text{H}$) absorption \citep{duccio}.  Moreover, the prevalence of FRII-LERGs increases when examining samples of RGs with lower radio powers, typically in the millijansky regime \citep{FRIIcat, Mingo}.
Low-power RGs with radiatively efficient engines have also been observed \citep{gendre}. The well-known RG 3C~120, morphologically classified as an FRI \citep{WA87} but with broad and intense optical lines \citep{B87}, is a typical example.

The observation of a significant number of RGs belonging to transitional classes is opening a new debate on the relation between accretion and ejection.
Based on their X-ray results, \cite{duccio}, for example,  proposed two possible explanations for the nature of FRII-LERGs: (i) FRII-LERGs represent an evolved class of FRII-HERGs (the evolutionary scenario) and (ii) jets in LERGs are initially similar, with the environment shaping their kiloparsec-scale morphology (the environmental scenario).
In the first hypothesis, all FRII galaxies originate as HERGs. When their fuel is depleted, the accretion disk transitions from an SS-disk to a radiative inefficient accretion flow state. The observation of FRII-LERGs is thus attributed to a temporal disconnection between the radio-extended structure, formed during the earlier active phase, and the nuclear region, which, having recently changed its state, alters the ionization state of the narrow-line region.
 In the second hypothesis, inefficient accretion flows give rise to only one kind of jet, while interaction with the surrounding medium primarily modifies its morphology on kiloparsec scales.

A third hypothesis, i.e., the engine scenario, emerged from the study of 406 RGs \citep{grandi21}  extracted from a sample constructed by cross-correlating Data Release 7 of  Sloan Digital Sky Survey and the Faint Images of the Radio Sky at Twenty-cm (FIRST) survey \citep{best12}. By examining faint objects with well-defined radio and optical classifications down to millijansky flux densities, \cite{grandi21} showed that sources with similar inefficient accretion rates can exhibit different radio luminosities and morphologies, ranging from compact \citep{FR0Cat} to FRI \citep{FRIcat} and FRII \citep{FRIIcat}.
This finding suggests that the nature of the jets  may depend not solely on the accretion regime but also on other nuclear properties, such as the black hole's mass, spin, and magnetic field.

To shed light on these possible scenarios, we focused on the milliarcsecond radio properties of the 3CR RGs at redshifts z$<0.3$ studied by  \cite{buttiglione09,buttiglione10,buttiglione11}, as inferred from very long baseline interferometry (VLBI) observations.
We used this sample to compare FRII-LERGs and FRII-HERGs on parsec scales. We studied FRIs in order to derive more general conclusions.  
The cut at jansky flux densities of the 3CR catalog prevents the comparison of FRI and FRII sources in similar redshift ranges. 
This is a selection effect that several authors have discussed. For example,
 \cite{giovannini88,g05}, studying the correlation between core and total radio power in a large sample that included low-luminosity B2 \citep{B2} and high-luminosity 3CR sources, found no significant evolutionary effects up to z$<0.3$.

Moreover, the most recent millijansky study of \cite{grandi21} confirms that the radio luminosity difference between FRI-LERGs and FRII-LERGs persists even when the two classes are matched in redshift.
However, our investigation primarily focuses on FRIIs
with different accretion regimes, for which the redshift ranges are comparable.

Adopting the same cosmology as \cite{buttiglione09}, i.e., $H_0 = 71$ km s$^{-1}$ Mpc$^{-1}$, $\Omega_m = 0.3$, and $\Omega_\Lambda = 0.7$, we explored and compared their radio and optical luminosities on parsec scales, noting that for z=0.3, 1 mas corresponds to 4.4 pc.
In particular, we aim to investigate whether the separation among radio-optical classes is already evident at early ejection stages before the jets interact with the surrounding kiloparsec-scale medium.

\section{The 3CR-VLBI samples}

We studied the 3CR RGs at z$<0.3$ that were optically classified by \citet[hereafter B10]{buttiglione09,buttiglione10,buttiglione11} and have a defined radio classification. The sample consists of 79 objects. We checked how many of these sources are present in the Fundamental Radio Catalog (rfc$\_$2024b version) of compact radio sources \citep{Popkov}, which 
contains positions and fluxes for 21,907 compact radio sources derived from the reanalysis of 1,088 VLBI experiments. For each object, the catalog provides the precise positions and the total ($f_{tot}$) and unresolved ($f_{un}$)  flux densities\footnote{The unresolved flux density refers to the source component not resolved by the VLBI.} at 2.2-2.4 GHz (S band), 4.1-5 GHz (C band), 7.3-8.8 GHz (X band), 15.2-15.5 GHz (U band), and 22-24.2 GHz (K band). 

The 3CR and Fundamental Radio Catalogs were cross-correlated using a radius of 5 arcminutes searching for observations in at least one of the four bands. The Very Large Array archive images\footnote{http://cutouts.cirada.ca/} of the RGs in the cross-correlated output list were then inspected to verify the reliability of the associations. The final number of confirmed associations was 55.

 The VLBI luminosities were calculated in the X band ($\sim 8$ GHz),  encompassing the most significant number of observed RGs (52 out of 55). For sources without observations in the X band, the total and unresolved flux densities at 8 GHz were estimated from the nearest available band assuming $F(\nu) = k\nu^\alpha$ with $\alpha=0$. The unresolved flux densities were available for 44 objects. The luminosities are K-corrected. The final 3CR-VLBI sample is shown in Appendix A. 
 
\begin{figure*}[htbp]
    \includegraphics[width=1.\textwidth]{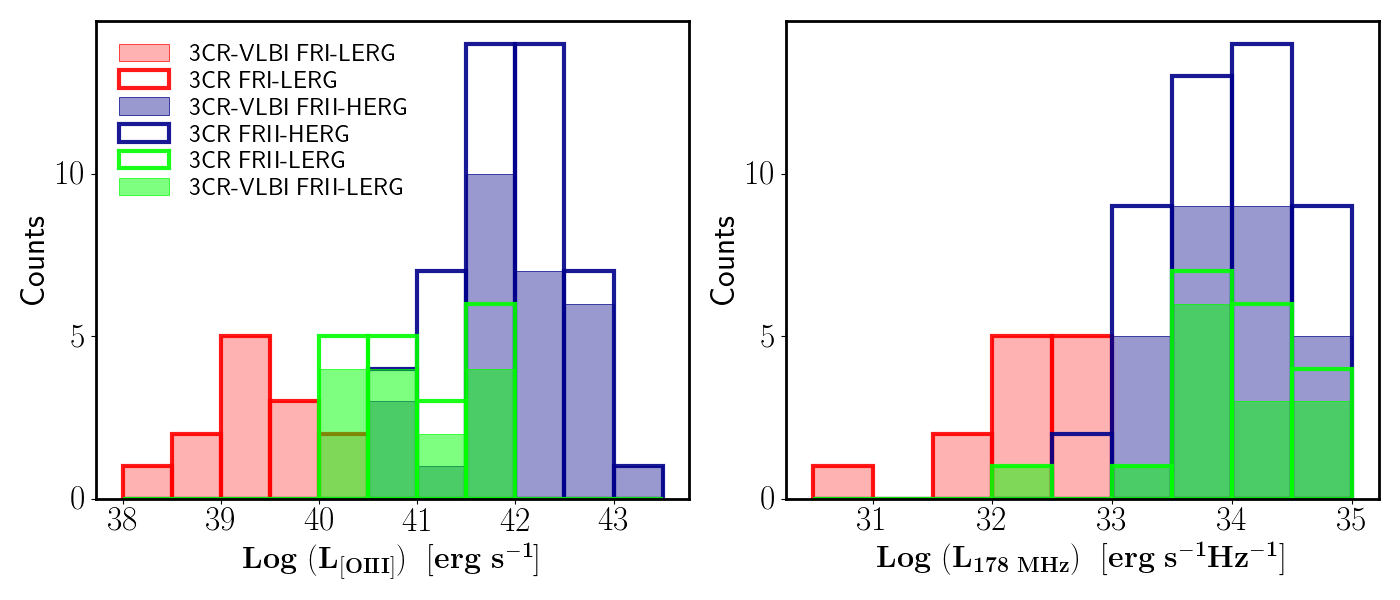}
\caption{Luminosity distributions of the [O III] $\lambda5007$ optical lines  ({\it left panel}) and at 178 MHz ({\it right panel}) for different radio-optical classes from the original 3CR sample studied by \cite{buttiglione09,buttiglione10,buttiglione11}and its subset with milliarcsecond observations (3CR-VLBI sample).
The filled portion of each histogram bin represents objects with VLBI observations, while the total height of the bin corresponds to the number of sources in the original 3CR sample. The distribution of the VLBI subsample closely follows that of the original sample.}

\label{figura1a}
\end{figure*}
The classes from B10 are well represented in the 3CR-VLBI subset.
Figure \ref{figura1a} illustrates the 3CR distribution of the [OIII]$\lambda5007$ optical emission line luminosity ({\it left panel}) and the radio luminosity at 178 MHz ({\it right panel}). The filled portion of each histogram represents the VLBI subsample, which closely mirrors the B10 sample. Since $L_{178~\mathrm{MHz}}$ serves as a proxy for the kinetic power of jets \citep{cg08} and $L_{[OIII]}$ reflects the accretion rate \citep{Heckman04}, the VLBI subsample provides a good representation of the different nuclear engines and extended kiloparsec-scale morphologies of the initial sample.

Following the study of \cite{rfc}, we considered the parsec compactness parameter at 8 GHz ($C_{8~GHz}$ =f$_{un}$/f$_{tot}$).  A value of $C_{8~GHz}=1$ corresponds to a compact (unresolved) source at parsec scales, while lower values indicate the presence of extended structures. It is analogous to the traditional ratio that roughly estimates jet orientation by comparing core and extended emission on kiloparsec scales. $C_{8~GHz}$ is useful for identifying potential observational biases in our VLBI sample. For instance, a narrow parsec-scale compactness distribution within a radio-optical class would suggest poor sampling, with RGs covering only a small range of angles. Similarly, if FRIIs were mostly unresolved due to their larger distance, their compactness values would be tightly clustered around $C = 1$. As shown in Fig. \ref{figura1b} ({\it left panel}), potential selection effects are negligible, as the parsec-scale compactness is similarly distributed across the analyzed radio-optical classes and spans a large range.

We applied the Mann-Whitney (MW) and Kolmogorov-Smirnov (KS) tests to assess potential differences between samples statistically. Both tests are nonparametric and compute the probability ($p_0$) that two samples derive from the same distribution (null hypothesis H$_0$).
The KS test is sensitive to differences across the entire distribution, whereas the MW test, which handles tied values better, is more sensitive to differences in medians. 
We assumed that two datasets are different if $p_0 < 5\times 10^{-2}$ in both tests.
The p-values were above the adopted thresholds for all class pairs examined, preventing us from rejecting the null hypothesis.

\section{Jet comparison at parsec and kiloparsec scales}
The luminosity at 178 MHz provides information about the kiloparsec-scale structures dominated by steep-spectrum lobes. In contrast, the luminosity at 8 GHz offers the opportunity to explore the jets on parsec scales. Indeed,
the milliarcsecond beam sizes of the maps in our sample correspond mostly (75\%) to linear dimensions, uncorrected for projection effects, smaller than 10 pc. For the rest, the linear extension is tens of pc, and only one case reaches 260 pc.

\begin{figure*}[htbp]
    \includegraphics[width=1.\textwidth]{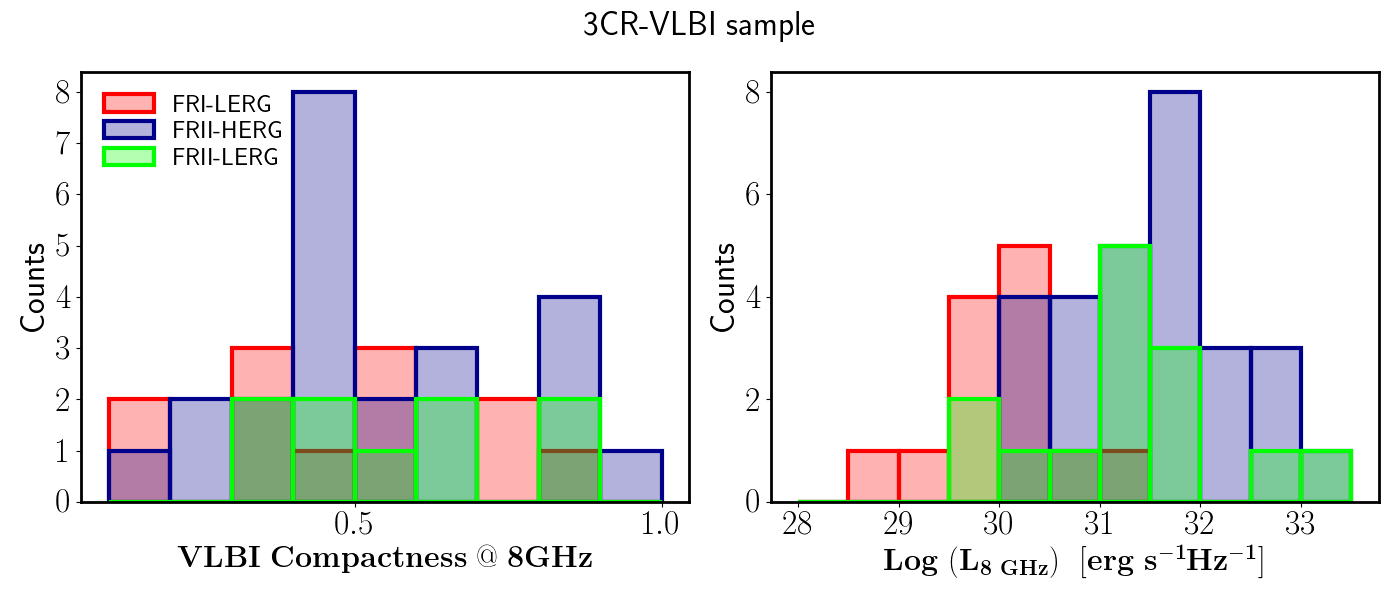}
\caption{ Compactness $C_{8~GHz}$ =$f_{un}/f_{tot}$ ({\it left  panel}) and luminosity ({\it right panel})  distributions  at 8 GHz of the 3CR-VLBI RGs. The bin size for each plotted luminosity distribution is 0.5. The compactness bin size is 0.1.}
\label{figura1b}
\end{figure*}

 In Fig. \ref{figura1b} ({\it right panel}), the luminosity distribution of each VLBI radio-optical class at 8 GHz is shown.
Interestingly, the radio-optical classes have the same relative arrangement as in the corresponding 178 MHz distribution represented by the filled bins in Fig. \ref{figura1a} ({\it right panel}). FRIs and FRIIs are separated into distinct classes, while FRII-LERG and FRII-HERG sources exhibit comparable luminosities.
The larger overlap among the classes observed at 8 GHz is expected, as moderate (the sample does not include blazars) relativistic effects can amplify the luminosity scatter. This beaming effect, already pointed out by \cite{giovannini88,giovannini2001}, is quantified by the well-known $P_{core}-P_{tot}$ correlation.

We find no statistical evidence of a luminosity offset between FRII-LERGs and FRII-HERGs with $p_{0}^{MW}=0.3$, and $p_{0}^{KS}=0.4$. We note that no redshift-related biases are present.

\begin{figure}[t]
\includegraphics[width=10.0cm,height=8cm]{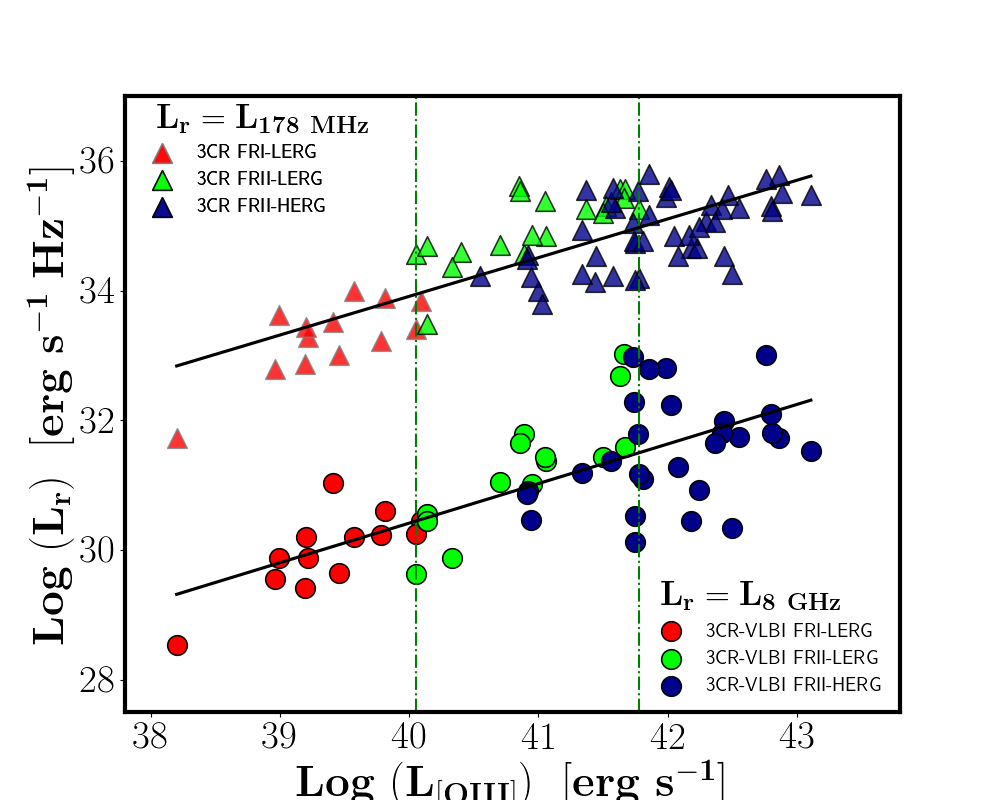}
\caption{
Radio luminosity ($L_{\rm r}$) versus [OIII] luminosity for the different radio-optical classes. Triangles correspond to the B10 sample, and circles the 3CR-VLBI subsample. The 3CR radio luminosities are measured at 178~MHz, whereas the 3CR-VLBI luminosities are at 8~GHz. The 178~MHz luminosities have been shifted by one decade (+1) to clarify the comparison between low and high frequencies. The vertical dashed-dotted green lines indicate the lower and upper [OIII] luminosity values for the 3CR FRII-LERGs, highlighting how the separation of RGs into different radio-optical classes is notably similar at 178~MHz and 8~GHz.}
\label{figura2}
\end{figure}

In Fig. \ref{figura2}, the luminosities at 8 GHz and 178 MHz are plotted as a function of the [OIII] line luminosity. No substantial difference is observed at the two frequencies. 
When plotted as a function of the accretion rates, the radio-optical classes maintain the same relative position, whether considering the luminosity at 8 GHz or 178 MHz.
FRIs are less luminous in radio and have weaker optical nuclei. FRII-HERGs are clustered in the right region of the plot, having high radio and [OIII] luminosities. FRII-LERGs are in between the two extremes.
 Figure \ref{figura2} unambiguously shows that FRII RGs with different accretion rates already have comparable radio luminosity on parsec scales.

\section{Discussion}

  This study explores the properties of 3CR RGs observed with VLBI at milliarcsecond resolution. While similar analyses exist in the literature \citep{G94, Cotton, Venturi, Lara, giovannini2001, g05, Liuzzo}, this work offers a novel approach by examining a range of radio morphologies alongside diverse accretion regimes. Specifically, we used the [OIII] luminosity as a proxy for accretion and compared luminosities at 8~GHz and 178~MHz.

Our analysis reveals that: 

\begin{itemize}
    \item  The different optical-radio classes occupy the same relative positions in the L$_{OIII}-L_{radio}$ plot, whether considering fluxes from extended kiloparsec-scale radio structures or emission from parsec-scale regions.\item Although FRII-LERGs and FRII-HERGs are associated with different accretion regimes, at each radio frequency, they exhibit a comparable range of radio luminosities.

\end{itemize}

\noindent These findings strongly weaken the idea that FRII-LERGs are evolved HERGs that have switched off due to fuel depletion.
That hypothesis relies on a temporal disconnection between the radiation from the narrow-line regions ($L_\text{[OIII]}$), located tens to hundreds of light-years from the central engine, and the radiation observed at 178~MHz, which mainly originates from $10^{8}$ year old structures.
The observation of powerful jets in LERGs at parsec scales challenges the evolutionary scenario. 

The results at 8~GHz do not even support the environmental scenario, which attributes the different radio morphologies primarily to the interaction of the jet with the kiloparsec-scale medium.
We note that milliarcsecond-scale observations map regions very close to the Bondi radius. For 19 RGs in our sample, \cite{Balmaverde08} provided estimates of $R_\text{Bondi}$ that align well with the VLBI resolution.
Although still debated \citep{bia}, this radius is thought to coincide with the zone where the jet's expansion changes from parabolic to conical, corresponding to an evolution from a magnetically dominated phase to one of free expansion \citep{kovalev}.
It seems more plausible to hypothesize that the differentiation  among various types of RGs occurs during the initial ejection phase when the jet is still accelerating and interacting with the nuclear environment.

 \cite{bia} discusses a possible link between the accretion mode and the VLBI properties of the jets.
  Their analysis of mass-scaled jet expansion profiles from 13 RGs revealed that both FRII-HERGs and FRI-LERGs can produce jets surrounded by external layers. However, the base of the expelled plasma in radiatively efficient disks appears to be more extended, with an outer launch radius exceeding even 100 Schwarzschild radii.
Theoretically, two mechanisms have been proposed to explain jet production: the Blandford-Znajek (BZ) model \citep{BZ} and the Blandford-Payne (BP) model \citep{BP}. The BZ model suggests that the jet is formed by extracting rotational energy from a supermassive, rotating black hole. The jet power is proportional to $P_{BZ} \propto \Phi^2 a^2 M_{BH}^2$, where $a$ is the black hole spin and $\Phi$ is the magnetic field at its horizon. The BP mechanism involves rotating magnetic field lines anchored in the accreting material, which centrifugally drive outflows of matter from a magnetized disk. 
Thus, the differing jet topologies could suggest that, while in LERGs the BZ mechanism is primarily responsible for jet ejection, in HERGs a BP contribution from the disk is also at play. 

Within this general picture, the observation of FRII-LERGs could indicate that the passage between efficient and inefficient regimes is less sharp than suggested by the optical LERG and HERG classification. Theoretical studies indicate that inefficient accretion configurations involving a combination of a hot flow and a cold disk can exist (see Fig. 1 of \citealt{yuan}). 
Another option, which does not necessarily conflict with the previous one, assigns a key role to the black hole and its immediate environment. The acceleration imparted to jets in any accretion regime would depend on the properties of the black hole, such as its mass, spin, and the magnetic field accumulated at its horizon (the engine scenario). 

Recent simulations seem to support this view.
\cite{Rossi} performed three-dimensional numerical simulations of magnetized relativistic jets propagating in a uniform-density environment. By considering jets with low densities (i.e., low power) and different magnetizations, they found that different types of jets can be produced. Jets with a higher magnetization are well collimated and maintain a high Lorentz factor, whereas jets with a lower magnetization decelerate within approximately 1~kpc, appearing as FRI jets on larger scales.
Via general relativistic magnetohydrodynamics simulations, \cite{lalakos} found that a rapidly rotating black hole accreting weakly magnetized gas from outside the Bondi radius can go through a phase in which a magnetically arrested disk is formed. 
In the magnetically arrested disk configuration, the magnetic forces are in equilibrium with the ram pressure from the accreting gas, allowing the black hole to maintain the maximum amount of magnetic flux at its event horizon \citep{mad1,mad2}. In such a configuration, both FRI and FRII jets can be launched.

In conclusion, the radio morphology of RGs on kiloparsec scales appears to be strongly influenced by processes occurring during the initial plasma ejection phase. The type of accretion disk (along with its associated winds) and the properties of the black hole are likely the primary drivers of their segregation into optical-radio classes.
\begin{acknowledgements}
We thank the editor for the support during the revision process.
We would also like to thank  Dr. Leonid Petrov for granting us permission to use the data from the Radio Fundamental Catalog, available at DOI: 10.25966/dhrk-zh08. 
We express our sincere gratitude to Dr. Rocco Lico for his help in utilizing the Radio Fundamental Catalog.
This research has made use of the CIRADA cutout service at URL cutouts.cirada.ca, operated by the Canadian Initiative for Radio Astronomy Data Analysis (CIRADA). CIRADA is funded by a grant from the Canada Foundation for Innovation 2017 Innovation Fund (Project 35999), as well as by the Provinces of Ontario, British Columbia, Alberta, Manitoba and Quebec, in collaboration with the National Research Council of Canada, the US National Radio Astronomy Observatory and Australia’s Commonwealth Scientific and Industrial Research Organisation. This research has used the NASA/IPAC Extragalactic Database (NED), which is operated by the Jet Propulsion Laboratory, California Institute of Technology, under contract with the National Aeronautics and Space Administration.
\end{acknowledgements}
\bibliography{3CR_VLBI}
\bibliographystyle{aa}

\onecolumn
\begin{appendix} 

\section{3CR-VLBI sample}
\small
  \begin{longtable}{lcllcccccc}

\caption{Properties of the 3CR-VLBI sample. }\\

\hline

Source & z & \multicolumn{2}{c}{CLASS} &Log(L$_{[OIII]}$) & Log(L$_\text{178 MHz}$) & Log(L$_\text{8 GHz}$) & C$_{8~GHz}$&\multicolumn{2}{c}{beam} \\

   &  & Radio      & Optical &(erg s$^{-1}$) & (erg s$^{-1}$ $Hz^{-1}$) & (erg s$^{-1}$ $Hz^{-1}$)&  &
  \multicolumn{2}{c}{mas x mas} \\
  \hline
  \endfirsthead
\hline
Source  & z & \multicolumn{2}{c}{CLASS} &Log(L$_{[OIII]}$) & Log(L$_\text{178 MHz}$) & Log(L$_\text{8 GHz}$) & C$_{8~GHz}$&\multicolumn{2}{c}{beam} \\ 

   &  & Radio      & Otical &(erg s$^{-1}$) & (erg s$^{-1}$ $Hz^{-1}$) & (erg s$^{-1}$ $Hz^{-1}$)&  &
  \multicolumn{2}{c}{mas x mas} \\

  \hline
  \endhead
  
  \hline
  \endfoot
  
  \hline
  \endlastfoot
  \\
    3C017 & 0.220 & FRII&   HERG & 41.99 & 34.44 & 32.80 &   0.46 &  1.97 &  0.82 \\
  3C018 & 0.188 & FRII &   HERG & 42.55 & 34.27 & 31.74 &   0.94 &  3.07 &  1.06 \\
  3C029 & 0.045 & FRI&   LERG& 40.09 & 32.84 & 30.44 &   0.81 &  2.75 &  1.54 \\
  3C031 & 0.017 & FRI &   LERG& 39.46 & 32.01 & 29.64 &   0.57 &  2.76 &  2.04 \\
  3C033 & 0.059 & FRII &   HERG & 42.18 & 33.65 & 30.45 &   0.87 & 11.52 &  0.85 \\
  3C040 & 0.018 & FRI &   LERG& 39.22 & 32.29 & 29.87 &   0.71 &  2.47 &  0.90 \\
 3C066B & 0.021 & FRI &   LERG& 40.05 & 32.40 & 30.25 &   0.52 &  1.50 &  0.94 \\
  3C078 & 0.030 & FRI &   LERG& 39.41 & 32.51 & 31.03 &   0.35 &  1.89 &  0.77 \\
  3C079 & 0.256 & FRII &   HERG & 42.86 & 34.78 & 31.73 &   0.68 &  3.20 &  1.41 \\
  3C088 & 0.030 & FRII &   LERG& 40.14 & 32.49 & 30.56 &   0.48 &  2.08 &  0.87 \\
  3C111 & 0.050 & FRII &   HERG & 42.44 & 33.54 & 31.99 &   0.29 &  3.48 &  1.64 \\
  3C133 & 0.278 & FRII &   HERG & 42.76 & 34.72 & 33.00 &   0.48 &  2.36 &  1.03 \\
  3C153\textsuperscript{$\sharp$}& 0.277 & FRII &   LERG& 41.63 & 34.56 & 32.68 & ... & 21.90 & 18.34 \\
  3C165 & 0.296 & FRII &   LERG& 41.67 & 34.57 & 31.59 &  ...& 59.30 &  1.84 \\
  3C166 & 0.245 & FRII &   LERG& 41.66 & 34.42 & 33.02 &   0.67 &  2.23 &  0.98 \\
3C197.1 & 0.128 & FRII &   HERG & 40.92 & 33.55 & 30.91 &   0.88 &  2.46 &  0.94 \\
3C213.1 & 0.194 & FRII &   LERG& 41.06 & 33.84 & 31.37 &   0.84 &  2.42 &  1.04 \\
  3C219 & 0.175 & FRII &   HERG & 41.77 & 34.53 & 31.79 &   0.56 &  2.49 &  0.77 \\
  3C227 & 0.086 & FRII &   HERG & 41.75 & 33.74 & 30.52 &   0.65 &  5.81 &  3.25 \\
  3C234 & 0.185 & FRII &   HERG & 43.11 & 34.47 & 31.52 &   0.43 &  6.43 &  2.67 \\
  3C236 & 0.099 & FRII &   LERG& 40.89 & 33.56 & 31.78 &   0.39 &  2.08 &  1.03 \\
  3C264 & 0.022 & FRI &   LERG& 39.20 & 32.43 & 30.20 &   0.59 &  2.33 &  1.01 \\
  3C270 & 0.007 & FRI &   LERG& 38.96 & 31.79 & 29.55 &   0.12 &  2.00 &  0.87 \\
3C272.1 & 0.003 & FRI &   LERG& 38.20 & 30.72 & 28.53 &   0.71 &  2.10 &  0.91 \\
  3C274 & 0.004 & FRI &   LERG& 38.99 & 32.63 & 29.87 &   0.17 &  1.22 &  0.65 \\
3C277.3 & 0.086 & FRII &   HERG & 40.94 & 33.21 & 30.46 &  ... &  7.28 &  2.57 \\
3C287.1 & 0.216 & FRII &   HERG & 41.73 & 34.04 & 32.97 &   0.46 &  2.78 &  1.03 \\
  3C288 & 0.246 & FRII &   LERG& 40.86 & 34.53 & 31.65 &   0.80 &  9.78 &  2.55 \\
  3C296 & 0.025 & FRI &   LERG& 39.78 & 32.22 & 30.23 &   0.46 &  2.42 &  0.84 \\
  3C303 & 0.141 & FRII &   HERG & 41.74 & 33.77 & 32.28 &   0.54 &  1.28 &  0.96 \\
3C303.1 & 0.267 & FRII &   HERG & 42.42 & 34.25 & 31.80 &  ... & 18.70 &  7.33 \\
  3C310$^{\sharp}$ & 0.054 & FRII &   LERG& 40.05 & 33.56 & 29.63 &  ... & 41.97 &  4.22 \\
  3C321 & 0.096 & FRII &   HERG & 40.91 & 33.49 & 30.86 &  ... &  2.68 &  1.26 \\
3C323.1 & 0.264 & FRII &   HERG & 42.80 & 34.31 & 32.10 &   0.87 &  2.38 &  0.91 \\
  3C327 & 0.104 & FRII &   HERG & 42.24 & 33.98 & 30.92 &   0.36 &  4.05 &  1.38 \\
  3C332 & 0.151 & FRII &   HERG & 41.81 & 33.77 & 31.09 &   0.50 &  2.73 &  0.96 \\
  3C338 & 0.031 & FRI &   LERG& 39.57 & 32.99 & 30.20 &   0.36 &  1.54 &  1.05 \\
  3C349 & 0.205 & FRII &   LERG& 41.50 & 34.20 & 31.43 &   0.53 &  3.20 &  2.17 \\
  3C353 & 0.030 & FRII &   LERG& 40.14 & 33.69 & 30.44 &   0.32 &  3.63 &  1.82 \\
  3C357 & 0.166 & FRII &   LERG& 40.95 & 33.86 & 31.02 &  ... &  4.30 &  2.70 \\
  3C381 & 0.161 & FRII &   HERG & 42.37 & 34.06 & 31.65 &   0.86 &  2.78 &  0.95  \\
  3C382 & 0.058 & FRII &   HERG & 41.78 & 33.19 & 31.17 &   0.27 &  2.50 &  1.12 \\
  3C388 & 0.091 & FRII &   LERG& 40.70 & 33.70 & 31.05 &   0.47 &  1.51 &  0.98 \\
3C390.3 & 0.056 & FRII &   HERG & 42.08 & 33.54 & 31.28 &   0.45 &  3.95 &  1.33 \\
  3C401 & 0.201 & FRII &   LERG& 41.05 & 34.38 & 31.43 &   0.65 &  3.04 &  1.02 \\
  3C403$^{\sharp}$ & 0.058 & FRII &   HERG & 41.75 & 33.16 & 30.13 &  1.00 & 19.42 &  6.77 \\
  3C410 & 0.289 & FRII &   HERG & 41.86 & 34.80 & 32.79 &   0.44 &  1.69 &  0.96 \\
  3C430 & 0.056 & FRII &   LERG& 40.33 & 33.36 & 29.88 &  ... &  5.39 &  3.01 \\
  3C436 & 0.215 & FRII &   HERG & 41.56 & 34.37 & 31.38 &  ... &  2.99 &  1.37 \\
  3C445 & 0.056 & FRII &   HERG & 42.50 & 33.26 & 30.33 &   0.39 &  7.36 &  1.40 \\
  3C449 & 0.017 & FRI &   LERG& 39.19 & 31.87 & 29.41 &  ...
  &  3.77 &  2.49 \\
  3C452 & 0.081 & FRII &   HERG & 41.34 & 33.94 & 31.19 &   0.14 &  2.77 &  1.15 \\
  3C456 & 0.233 & FRII &   HERG & 42.81 & 34.23 & 31.81 &   0.67 &  2.38 &  0.92 \\
  &&&&&&&&&\\
  3C459 & 0.220 & FRII &   HERG & 42.03 & 34.55 & 32.23 &   0.40 &  3.31 &  0.96 \\
  3C465 & 0.030 & FRI &   LERG& 39.81 & 32.89 & 30.60 &   0.37 &  1.92 &  0.90 
 \label{tab:sample}

 \end{longtable} 
 \tablefoot{The optical and radio classifications, the [O\,\textsc{iii}] luminosities as well as the kiloparsec-scale 178\,MHz luminosities  are taken from \citep{buttiglione09,buttiglione10}. The VLBI parsec-scale luminosities at 8\,GHz and the relative source compactness are derived from the Fundamental Radio Catalog \citep{Popkov}. 
The last column reports the corresponding VLBI observation's beam resolution (in mas). 

$\sharp$ 8 GHz luminosity extrapolated from the nearest available VLBI band. The corresponding beam dimensions refer to the original map used to estimate $L_\text{8 GHz}$. When the unresolved component at 8 GHz is unavailable, the compactness value is replaced by an ellipsis.}


 \end{appendix}
\vspace{1cm}

\end{document}